\documentclass{article}

\usepackage{arxiv}

\usepackage[utf8]{inputenc} 
\usepackage[T1]{fontenc}    
\usepackage{hyperref}       
\usepackage{url}            
\usepackage{booktabs}       
\usepackage{amsfonts}       
\usepackage{nicefrac}       
\usepackage{microtype}      
\usepackage{lipsum}		
\usepackage{graphicx}
\usepackage{doi}
\usepackage{ulem}
\usepackage{amsmath}

\usepackage{orcidlink}
\usepackage[affil-it]{authblk}
\usepackage{placeins}

\usepackage[backend=biber, style=apa, natbib=true]{biblatex} 
\title{Behavioral Outcomes of Human Cognitive Security within an Integrative Modeling Framework}

\author[1]{Aaron R. Allred*\,\orcidlink{0000-0001-5241-2830}}
\author[1]{Erin E. Richardson\,\orcidlink{0009-0005-3839-4759}}
\author[2]{Sarah R. Bostrom\,\orcidlink{0000-0001-6527-5978}}
\author[3]{James Crum\,\orcidlink{0000-0001-7806-8572}}
\author[4]{\\Chad Tossell\,\orcidlink{0000-0003-1662-9308}}
\author[2]{Richard E. Niemeyer\,\orcidlink{0000-0002-1058-253X}}
\author[3,5]{Leanne Hirshfield\,\orcidlink{0000-0003-0111-6948}}
\author[1]{Allison P.A. Hayman\,\orcidlink{0000-0001-7808-8557}}

\affil[1]{Smead Department of Aerospace Engineering Sciences, University of Colorado, Boulder, CO, USA}
\affil[2]{United States Air Force Academy, USAFA, CO, USA}
\affil[3]{Institute of Cognitive Science, University of Colorado, Boulder, CO, USA}
\affil[4]{Department of Human Factors, Safety \& Social Sciences, Embry-Riddle Aeronautical University, Daytona Beach, FL, USA}
\affil[5]{Department of Computer Science, University of Colorado, Boulder, CO, USA}

\makeatletter
\renewcommand{\maketitle}{%
  \begin{center}
    {\LARGE \bfseries \@title \par}
    \vskip 1em
    {\large \@author \par}
    \vskip 1em
    {\@date \par}  
    \vskip 2em
  \end{center}
}
\makeatother

\renewcommand{\shorttitle}{\textit Cognitive Security}

\hypersetup{
pdftitle={Behavioral Outcomes of Human Cognitive Security within an
Integrative Modeling Framework},
pdfsubject={q-bio.NC, q-bio.QM},
pdfauthor={Aaron R.~Allred, Erin E.~Richardson, Sarah R.~Bostrom, James Crum, Chad Tossell, Richard E.~Niemeyer, Leanne Hirshfield, Allison A.P.~Hayman},
pdfkeywords={Decision Making, Cognition, Information Based Threats}
}

\begin{document}
\maketitle

\renewcommand{\thefootnote}{\fnsymbol{footnote}} 
\footnotetext[1]{\textit{Corresponding author:} \href{mailto:aaron.allred@colorado.edu}{aaron.allred@colorado.edu}}
\renewcommand{\thefootnote}{\arabic{footnote}}   

\begin{abstract}
	Human decision-making under uncertainty faces growing challenges from information-based threats that pose risks to human cognitive processes and behavior. Although their potential harm is widely acknowledged, there remains no well-defined construct for characterizing the degree to which information-based threats influence changes in human judgments and decision-making, impeding theoretical advancement, measurement, and effective countermeasure development. Here, we introduce a human cognitive security construct focused on linking information-based threats to observable outcomes to bridge field-level definitions with operational measures by drawing from core mechanisms related to information processing and decision-making. To connect the information environment to behavior, we develop an integrative modeling framework that unifies Bayesian inference with affect-modulated decision valuation, capturing how cognitive resource allocation and affective valuation shape three core behavioral outcomes: veracity discernment, task-oriented actions, and information sharing. Through computational simulations, we demonstrate that this framework explains canonical phenomena, including cognitive heuristics, the illusory truth effect (R²=0.86, validated against empirical data), and incongruent veracity discernment and sharing behavior. We propose empirically grounded behavioral outcome measures of cognitive security to guide future empirical examinations. Finally, we outline how environment-specific elements, characterized by data availability and ecological constraints, affect individuals' cognitive security and identify future research directions.
\end{abstract}


\section{Introduction}
Information-based threats, such as the presentation of erroneous information, pose substantial risks to societal, political, and economic stability, with potential consequences ranging from compromised individual decision-making to large-scale manipulation of democratic processes and national security vulnerabilities \citep{adams_why_2023, broda_misinformation_2024, bostrom_formal_2024}. To address these challenges and more, the field of `cognitive security’ has emerged \citep{ask_cognitive_2025}.  However, the field today still lacks a human-level construct that clearly defines what it means for an individual or team to be cognitively secure and, stemming from this understanding, a unifying framework explaining the influence of information-based threats on the construct of cognitive security. Without both, theoretical progress, operationalization for measurement, and subsequently, the design and evaluation of effective countermeasures cannot advance.

In many ways, these gaps reflect the fragmentation of relevant literature. Much empirical work on information-based threats has emphasized evaluating truth judgments, whether individuals judge information as true or false (primarily through veracity discernment tests) \citep{murphy_what_2023}, yet these studies have scarcely examined consequent behaviors \citep{allen_addressing_2025}. By contrast, research in engineering psychology and human factors engineering has developed process descriptions of attention, perception, judgment, and choice (reviewed in \citep{wickens_engineering_2015}) but has not examined the distinctive effects of erroneous information, despite recognition of this domain as a promising foundation for tackling these threats \citep{endsley_combating_2018, karwowski_grand_2025}. Further, a comprehensive framework or model has yet to be developed, despite many theoretically rich but fragmented efforts. The qualitative information-processing model (IPM) from human factors \citep{wickens_engineering_2015, wickens_information_2021} describes information flow toward choice but lacks parameterized relationships that generate quantitative predictions. Bayesian and network models have captured belief updating and polarization \citep{cook_rational_2016, zmigrod_misinformation_2023} but terminate at belief states rather than connecting to task-relevant behaviors influenced by affect and other descriptive decision-making elements. This separation has left the broader field without a cohesive understanding of how information-based threats influence cognitive security and has precluded the development of predictive frameworks for systematic intervention.

Current cognitive security efforts are commonly described, mostly in military contexts, as the protection of human cognitive processes and decision-making from adversarial manipulation and exploitation \citep{army_cyber_institute_army_2024, noauthor_nato_2024}. Other researchers describe it broadly as the capability to detect, control, and counter information-based harms \citep{grahn_cognitive_2024, janzen_cognitive_2022} or as applications extending to non-human systems \citep{andrade_cognitive_2019, casino_unveiling_2025}. Most recently, a field-level conceptual definition has been proposed: cognitive security is "the state of having trusted boundaries protecting cognitive assets against all forms of unauthorized influence or access" \citep{ask_cognitive_2025}. While this domain- and species-agnostic definition provides conceptual clarity at the macro-level (broad, general theory), it is necessary to translate this definition to specific human behavioral outcomes at the micro-level (measurable, operational outcomes). To address the need for a middle/meso-level (operationalizable to the micro-level) construct of \textit{human cognitive security}, we focus on the individual and team information processing involved in evaluating and integrating information under uncertainty. We propose the following theoretical construct within the broader field:

\textit{Cognitive security is the degree to which people leverage constructive, veridical information over other information to make truth-aligned judgments and decisions.}

This construct reflects the ability to both incorporate constructive, veridical information and resist the influence of other (e.g., erroneous, manipulative, or uninformative) information. Here, we qualify veridical information as constructive to distinguish it from truthful information that aims to mislead judgments or decisions (e.g., paltering), consistent with information-integrity principles emphasizing appropriate use rather than mere exclusion \citep{lundgren_defining_2019}. As such, the proposed construct encompasses both direct information-based threats and indirect effects, such as cases where information is accurately used but judgments or decisions are still compromised. It also spans the entire information-processing chain: from evaluating information, through probabilistic reasoning to form judgments, to making decisions. Finally, this proposed construct also provides the foundation for developing a modeling framework that can generate quantitative predictions about cognitive security across contexts. Consistent with this defined construct, we define cognitive attacks as information-based threats that attempt to reduce cognitive security by disrupting the information processing underlying judgments and decision-making \citep{claverie_cognitive_2022}. This includes the direct presentation of erroneous information as well as indirect manipulation of factors that influence human information processing, such as strategically altering perceived values associated with outcomes. The primary utility of this proposed construct is that it foregrounds behavioral outcomes that can be modeled computationally and observed empirically, addressing this gap within the current body of cognitive security research.


A modeling framework encompassing the information processing chain (and reflecting operational measures of this construct) should account for key behavioral phenomena while providing mechanistic explanations of how they arise. People exhibit truth bias, a tendency to accept incoming information as true \citep{fiedler_chapter_2012, levine_accuracy_1999, pantazi_power_2018}, and show illusory-truth effects whereby repetition increases perceived truth when recognized as repeated \citep{bacon_credibility_1979, hasher_frequency_1977, reber_effects_1999, udry_illusory_2024, unkelbach_truth_2019}. Perceived source credibility, shared worldview, in-group status, and authority shape acceptance and influence \citep{brinol_source_2009, mackie_processing_1990, nadarevic_perceived_2020, pennycook_lazy_2019, pennycook_psychology_2021}. Affect plays a central role, with reliance on affect attributed as a causal predictor of accepting erroneous information \citep{martel_reliance_2020} and higher emotionality associated with increased engagement with erroneous information \citep{horner_emotions_2021}. Although countermeasures such as warnings, deliberation prompts, and metacognitive cues can reduce some vulnerabilities \citep{bago_fake_2020, jalbert_only_2023, tanaka_beyond_2025}, their effects are uneven and context-dependent, so they have not been demonstrated to be comprehensive solutions to mitigate cognitive security concerns (for a comprehensive review, see \citep{allred_decision-making_2025}).

Critically, these collective tendencies appear to reflect shared mechanisms identified over five decades of judgment and decision-making (JDM) research rather than unique susceptibilities to erroneous information. Humans routinely rely on heuristics and intuitive thinking, producing deviations from normative expectations even under veridical information \citep{kahneman_prospect_1979, tversky_judgment_1974, tversky_advances_1992}. Affect is one of the most influential mediators \citep{forgas_mood_1995, greifeneder_when_2011, lerner_emotion_2015}, directly linked to choice evaluations \citep{lerner_emotion_2015, phelps_emotion_2014, schulreich_fear-induced_2020} which are goal-dependent \citep{molinaro_intrinsic_2023}. Social factors also influence decision quality, including biased information sampling \citep{stasser_pooling_1985} and source-based information weighting \citep{schobel_social_2016}. Metacognitive regulation and deliberative interventions can shift behavior toward more analytic responding, but their efficacy depends on context \citep{evans_dual-process_2013, mata_metacognitive_2013, de_neys_perspective_2017}. Supporting this convergence, sharing of erroneous information has been suggested to be mediated by affect and availability heuristics \citep{lu_heuristic_2024}, and intuitive thinking tendencies may increase susceptibility to erroneous information \citep{ecker_psychological_2022, pennycook_shifting_2021, pennycook_lazy_2019}. This convergence suggests that vulnerability to information-based threats arises from how general cognitive constraints interact with threat-based information environments.

Despite this convergence, critical explanatory gaps remain. Existing work has documented truth bias and the illusory truth effect but has not yet produced a formal account describing how they arise towards truth judgments. Similarly, the dissociation between veracity discernment and sharing behavior has been observed but not mechanistically explained. The field would benefit from a unified framework that connects how information is evaluated towards judgments, how those judgments inform actions, and finally, how information is propagated to others (communication behavior). Without such a framework, these gaps have limited both theoretical progress and the development of targeted interventions. 

To address this need, we develop an integrative modeling framework that synthesizes broader decision-making theory with findings on information-based threats, formalizing how latent cognitive processes produce relevant and observable behavioral outcomes. Motivated by the success of Bayesian approaches in perceptual and cognitive science that reconcile apparent suboptimalities as constrained solutions given ecological, computational, and energetic limitations \citep{gardner_optimality_2019, gigerenzer_heuristic_2011, hahn_unifying_2024, jones_bayesian_2011, martignon_naive_2003, martignon_categorization_2008, rahnev_suboptimality_2018, wei_bayesian_2015}, our framework models how limited computational resources, affective valuation, source credibility, and other factors jointly shape probabilistic judgments and choices.  We validate the framework through computational simulations grounded in empirical findings that span both domains, demonstrating that it captures phenomena while providing mechanistic insight into their origins. This harmonized approach enables the systematic study of context-specific factors affecting cognition and behavior to support the development of targeted interventions to protect judgments and decision-making.

\section{Results}
\label{sec:Results}

\subsection{A Framework Unifying Information Processing and Information-Based Threats}
\begin{figure}
	\centering
	\includegraphics[width=1\textwidth]{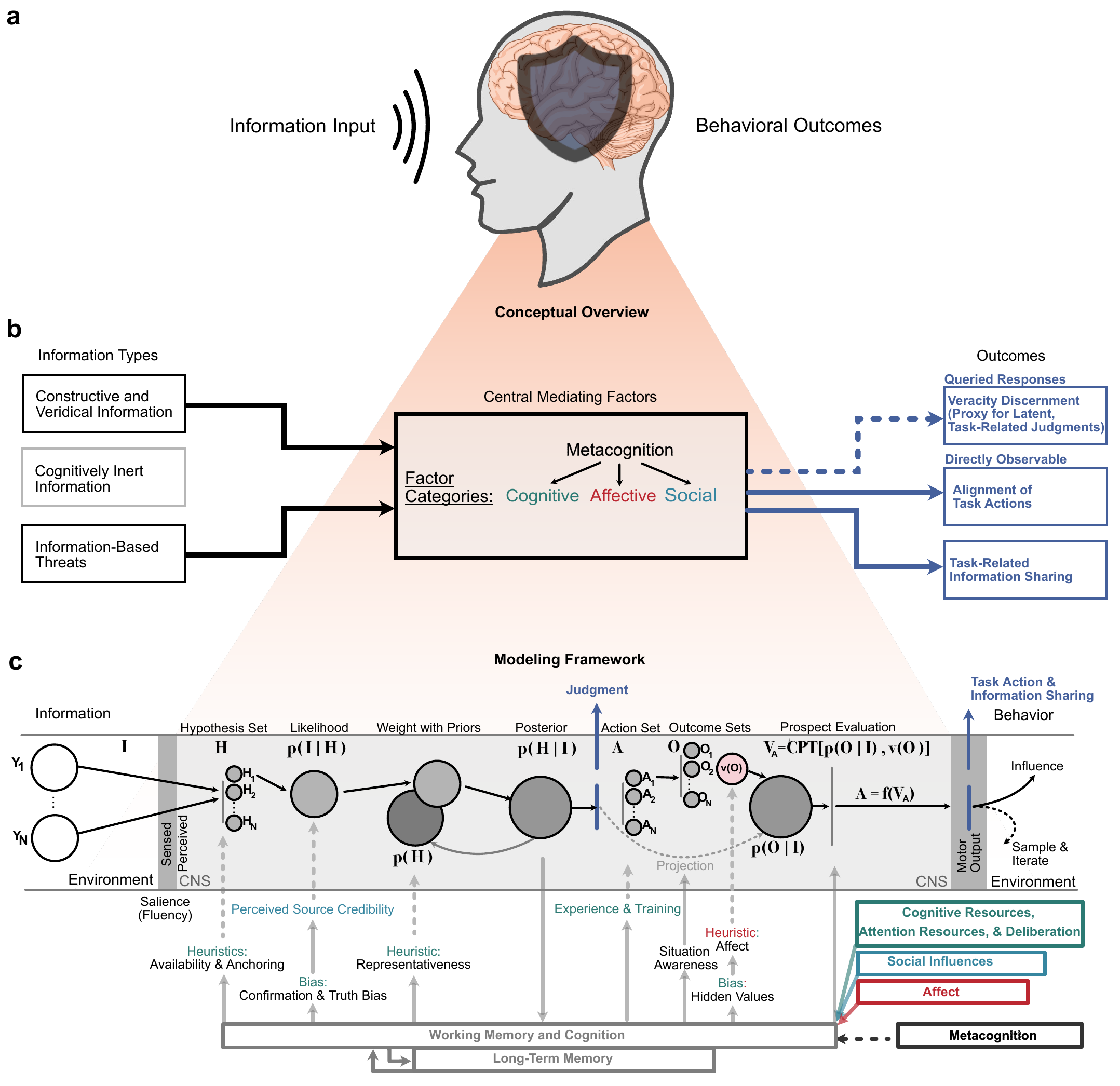}
	\caption{From conceptualization to the implementation of a modeling framework, we examine how information-based threats influence human judgment and decision-making under uncertainty toward behavioral outcomes of cognitive security. \textbf{a}. At the core of this study, we examine the influence of information processing on behavior. \textbf{b}. Centrally processed information is mediated by cognitive, affective, and social factors (all regulated by metacognition; hence the dashed line in panel c). Specifically, we model behavioral outputs: queried veracity discernment, directly observable task actions, and directly observable information sharing. \textbf{c}. From left to right, information in the environment must be sensed, perceived, and attended to in order to be processed by the central nervous system towards formulating judgments and decisions. Next, a set of potential hypotheses is considered, and each hypothesis is assigned an evidence-based evaluation, informed by sampled high-level information cues, an allocation of cognitive resources, and the perceived source credibility associated with each piece of information. Following posterior assessment, probabilities of various outcomes associated with a considered action set are computed, often requiring a transition from comprehension to projection. Qualitatively depicted here, judgments and decisions can be further influenced by metacognitive regulation and long-term memory (capturing expertise, training, and recall), which will further dictate the specific descriptive implementation of this modeling framework. Circles notionally represent distinct quantities as they progress through the information processing chain.}
	\label{fig:fig1}
\end{figure}

To address critical gaps in understanding how information-based threats shape human judgment and decision-making under uncertainty, we introduce a descriptive and parametric modeling framework (Figure \ref{fig:fig1})  derived from interdisciplinary research to reveal new mechanistic insights at the individual cognitive security level. As shown in Figure \ref{fig:fig1}b, the framework begins by parsing the available information in the environment and how the human processes it. We categorize these available information inputs as: (1) constructive and veridical information, (2) cognitively inert information, which is not meaningfully integrated into information processing, and thus is present but not impactful towards cognitive security, and (3) information-based threats, which threaten the use of constructive and veridical information towards judgments and decision-making. The next stage of the framework describes how humans process this information. Human information processing, mediated by cognitive, affective, and social factors, influences the ultimate outcome of the situation. To capture their interrelationships, this framework then yields the behavioral aspects of cognitive security as outcomes influenced by these mediators. Specifically, we capture the following behavioral aspects of cognitive security: (1) queried truth judgments, assessed using veracity discernment tasks (i.e., indirectly observed), (2)  directly observable task-oriented actions, and (3) directly observable information sharing.  

To formalize this framework, we develop a computational modeling framework (Figure \ref{fig:fig1}c), based on a two-stage process of decision-making \citep{mosier_judgment_2010}. On the front end, probabilistic assessments are generated via descriptive Bayesian inference, where information is evaluated through likelihood functions constrained by cognitive resource allocation across a considered hypothesis set. To model this relationship, we introduce a cognitive resource allocation function that quantifies contributions of uncertain information to these hypotheses (see Methods). Posterior assessments inform action values, which are evaluated through risk and benefit assessments based on Prospect Theory. Potential actions are constrained to a finite set, and outcome valuations are further modulated by affective and goal-driven parameters describing the value function.

\subsection{Broadly Capturing Judgments and Decision-Making}
\begin{figure}
	\centering
	\includegraphics[width=1\textwidth]{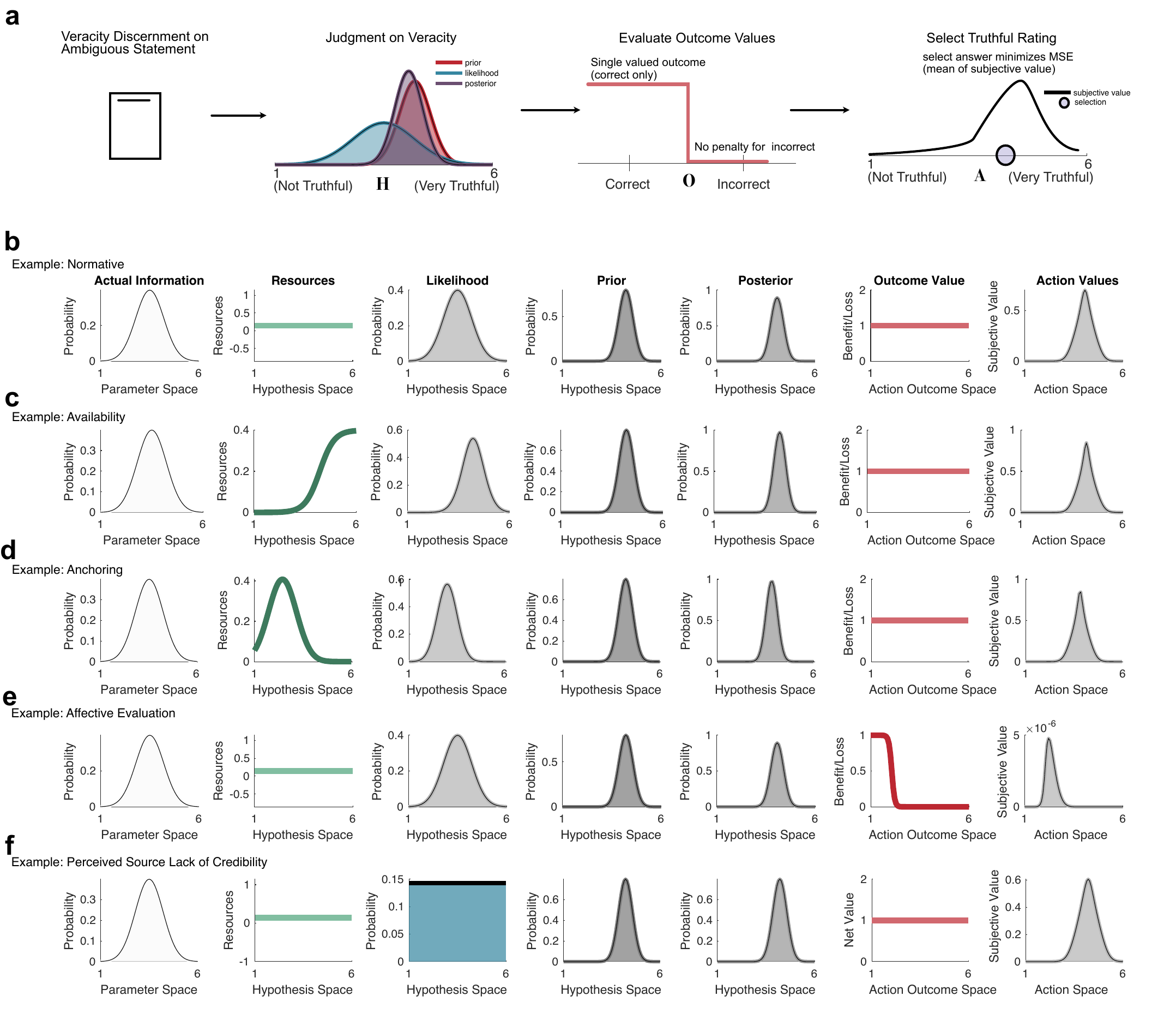}
	\caption{\textbf{a}. A model implementation of a veracity discernment task using the modeling framework presented in Figure \ref{fig:fig1}c. \textbf{b}. The normative case where an equal distribution of cognitive resources is allotted, producing a likelihood function that mirrors the underlying uncertainty in the information assessed via an ideal observer (see Methods for detailed implementation). Additionally, this model incorporates unbiased affective valuation of each outcome. \textbf{c-d}. Here, we explore how different cognitive resource mappings produced can describe the emergence of heuristics noted in formulating judgments under uncertainty: availability and anchoring, respectively. Here and throughout Figures \ref{fig:fig3} and \ref{fig:fig4}, saturated colors indicate where model inputs deviate from the normative case. \textbf{e}. Here, the effect of assigning affective values associated with each action being correct is explored. In this case, selecting 1 being correct is given a disproportional weighting, resulting in a shift in the selection on the veracity discernment task. \textbf{f}. Finally, a lack of perceived source credibility is modeled as a likelihood function that carries no information (as the information is disregarded on the basis of its credibility assessment), resulting in no update of the posterior probability density function from the prior.}
	\label{fig:fig2}
\end{figure}

With our modeling framework, we first examined how different allocations of cognitive resources, expressed through the cognitive resource mapping function on a hypothesis space, produce different judgments under information uncertainty during a veracity discernment task (depicted in Figure \ref{fig:fig2}a), an implementation aligned with measures in the literature. Here, the veracity discernment task is the evaluation of the truthfulness of a piece of information on a scale of not truthful to very truthful (i.e., H $\in$ [1,6]). In all cases, the amount of information contained is consistent. In the normative condition (Figure \ref{fig:fig2}b), the internal, centrally held likelihood function reflects the underlying uncertainty inherent to the provided statement (i.e., the uncertainty in the actual information content). From here, a novel piece of information is weighted with a uniform resource mapping and prior, and the posterior estimate is directly mapped to the probability of selecting each value along the continuous action space. However, across other simulations (Figure \ref{fig:fig2}c-d), the likelihood evaluation—and consequently the posterior probability distribution (thus the queried judgment) varies based on how cognitive resources are distributed across the hypothesis space. We demonstrate that distributions of cognitive resources produce two canonical cognitive heuristics: availability (Figure \ref{fig:fig2}c) and anchoring (Figure \ref{fig:fig2}d). For availability, biased resource allocation toward truthful hypotheses shifts the likelihood and posterior to be denser for truthful hypotheses upon which resources are prioritized. For anchoring, biased resource allocation centered on a specific hypothesis (for example, due to an external source such as a survey's slider location \citep{thomas_slider_2019} or an internal factor such as a prior selection) shifts the likelihood and posterior to be denser for the centered, "anchored," hypothesis upon which resources are prioritized.

Following this, we modeled how different allocations of values associated with each behavioral selection being correct (Figure \ref{fig:fig2}e) affect the queried selection, and how a discredited source can be modeled as a uniform likelihood, no longer influencing the truth judgment (Figure \ref{fig:fig2}d). In the case of altered value appraisals, the latent truth judgment is the same as the normative case, but the altered valuation changes the output veracity discernment behavior. For a discredited source, the uncertain information is attributed as uninformative, and the posterior remains the same as the prior. These simulations demonstrate the robustness of this framework when implemented during veracity discernment tasks, capturing how factors influencing information processing can influence the cognitive and behavioral aspects associated with cognitive security.

\subsection{Explaining Information-Based Threat Phenomena}
\begin{figure}
	\centering
	\includegraphics[width=1\textwidth]{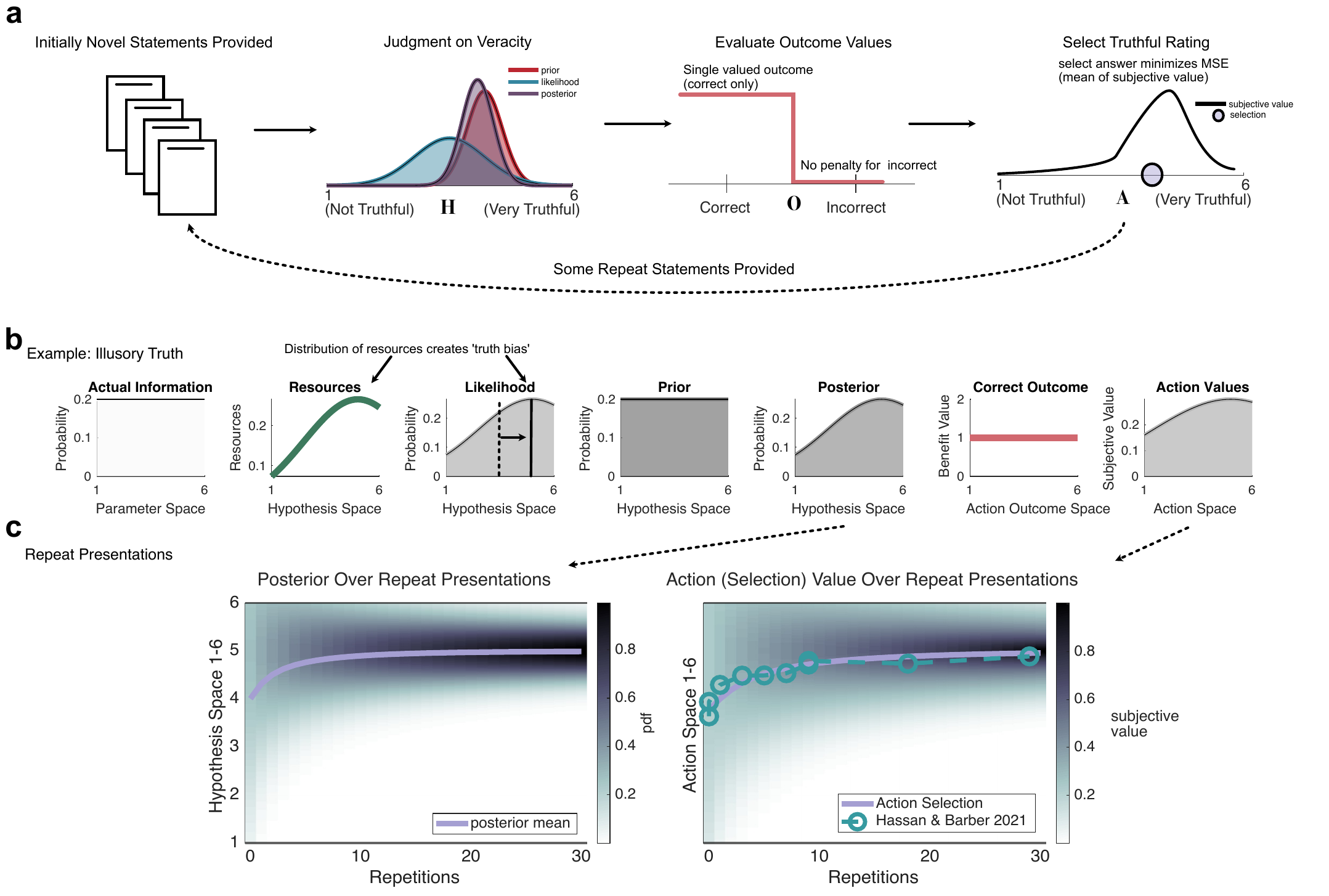}
	\caption{\textbf{a}. A model implementation (using the modeling framework presented in Figure \ref{fig:fig1}c) of a veracity discernment task over repeated exposure, producing the illusory truth effect. Computational simulations of the truth effect over information repetitions. \textbf{b}. Truth bias is modeled by biased resource mapping, producing a bias in the likelihood function. \textbf{c}. The emergence of illusory truth from truth bias in a likelihood function is demonstrated by plotting the mean of the prospective value (action selection) overlaid with empirical data \citep{hassan_effects_2021}. }
	\label{fig:fig3}
\end{figure}

Next, we asked to what extent this modeling framework can explain phenomena associated with information-based threats, including the illusory truth effect and misaligned veracity discernment and sharing behavior. In the illusory truth effect, repeated statements are evaluated to be more truthful \citep{hasher_frequency_1977,udry_illusory_2024}. To evaluate the illusory truth effect, we implemented a model (Figure \ref{fig:fig3}a) to replicate an empirical paradigm where participants were initially provided novel statements, some of which were repeated. This model implementation within our broader framework (Figure \ref{fig:fig1}c) sets the cognitive resource mapping to produce truth bias in the likelihood function (Figure \ref{fig:fig3}b), mirroring availability (Figure \ref{fig:fig2}c). The prior, which has a limited effect on the model output, is set to be naïve (uniform across hypotheses). Further details of this model implementation are outlined in the Methods. Our model outputs of veracity discernment behaviors (Figure \ref{fig:fig3}c) over repetitions (setting the prior to be the posterior of the previous repetition iteration) closely match the empirical finding (MSE=0.02 and $R^2$=0.86; with no prior distribution tuning) that subsequent repetitions of information produce the logarithmic true judgment behavior \citep{hassan_effects_2021}.

\begin{figure}
	\centering
	\includegraphics[width=1\textwidth]{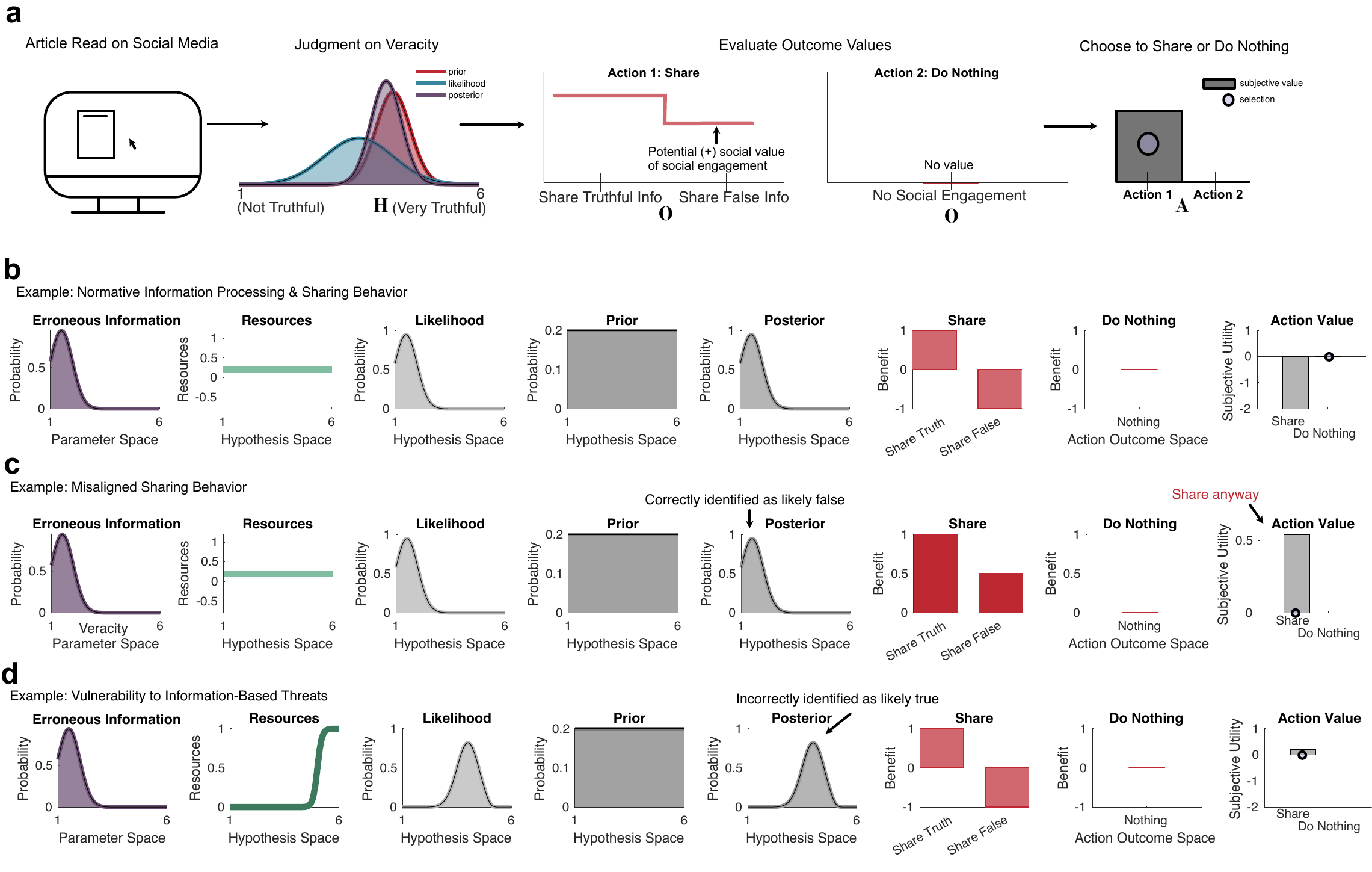}
	\caption{\textbf{a} A model implementation (using the modeling framework presented in Figure \ref{fig:fig1}c) of a share / no share paradigm (see Methods for the specific implementation). \textbf{b}. Normative sharing behavior, where erroneous information is correctly identified and not shared due to proportional benefits and losses associated with sharing veridical and erroneous information, respectively. \textbf{c}. Here, misaligned sharing behavior is described when the value of social engagement associated with sharing potentially false information outweighs the value of not sharing information. \textbf{d}. Full vulnerability to the information-based threat is realized as erroneous information is not correctly identified due to the allocation of cognitive resources along the hypothesis space, despite proportional benefits and losses associated with sharing veridical and erroneous information.}
	\label{fig:fig4}
\end{figure}

As a final evaluation, we asked how this modeling framework can describe misaligned veracity discernment and sharing behavior, a phenomenon where erroneous information is shared despite being correctly discerned as more likely to be false than true \citep{pennycook_shifting_2021,brashier_aging_2020}. To describe this behavior, we implemented a model (Figure \ref{fig:fig4}a) to produce this effect through an affective valuation. Here, our model indicates that this behavior will not emerge (Figure \ref{fig:fig4}b) when the loss associated with shared false information is equivalent to the benefit of sharing truthful information. However, our model indicates that this behavior will emerge (Figure \ref{fig:fig4}c) when the benefit of social engagement associated with sharing potentially false information outweighs the loss associated with not socially engaging. Since we set the value of not socially engaging (i.e., not sharing information) to be null, there must only be a positive value, no matter how small, associated with social engagement through sharing. 

\FloatBarrier

\subsection{Impact of the Information Environment on Cognitive Security}
\begin{figure}[!htbp]
	\centering
	\includegraphics[width=1\textwidth]{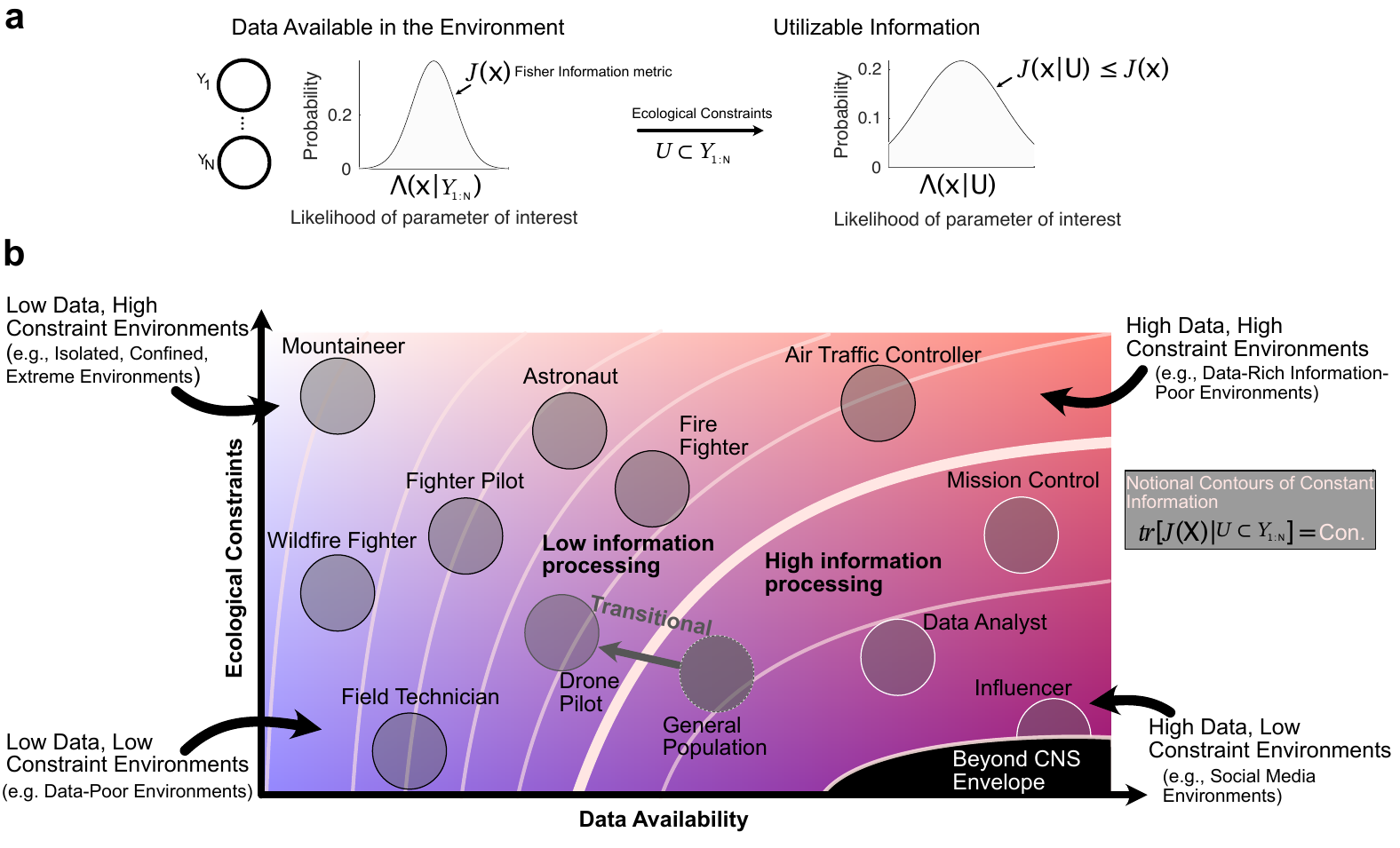}
	\caption{\textbf{a} A depiction of how ecological constraints, such as time pressure, can result in only a subset of information that can be processed by the human, denoted as utilizable information. \textbf{b}. The information processing spectrum of decision-making conceptually presented and categorized into low information processing (black outline) and high information processing environments (white outline), given shared elements expected to modulate judgments and decision-making, thus vulnerability to information-based threats. The information processing spectrum incorporates the amount of data available in the environment as well as ecological constraints on operators. Notional contour lines illustrate constant information across varying ecological constraints and data availability. On the most extreme end of low information processing environments (upper left corner), decreased information access and high ecological constraints largely coincide with operations in ICE environments. On the extreme end of high information processing environments (lower right corner), increased information access and low ecological constraints coincide with interactions in social media environments. Of importance for cognitive security are transitional groups (center; dashed outline), where information-based threats in one domain could spill over into impaired behavioral aspects in another domain. Notional naturalistic operational roles are assigned across the information processing spectrum. }
	\label{fig:fig5}
\end{figure}

Finally, our framework demonstrates that cognitive security is fundamentally shaped by the information environment, where both data availability and ecological constraints jointly influence how much information can be effectively used towards judgments and decision-making. Recognizing the environment is an important factor, we introduce an information process spectrum characterized by these two critical dimensions (Figure \ref{fig:fig5}).  Data availability ranges from information-poor (i.e., sparse) to information-rich, while ecological constraints determine how much available information can be practically used.  We use the trace of Fisher Information (tr(J)) as a quantitative metric for statistical or inferential environmental information content, useful within mathematical psychology \citep{ly_tutorial_2017}. As with Shannon’s information measure \citep{shannon_mathematical_1948}, Fisher information is a formal statistical, syntactic quantity describing parameter identifiability from data \citep{kolchinsky_semantic_2018}. While it gains semantic relevance when parameters correspond to states of interest in the world, the measure itself does not encode that meaning in isolation, and it must be interpreted within a domain context. Thus, the Fisher Information Matrix here quantifies the amount of information from available data ($Y_{1:N}$) that can be inferred about states of interest ($X \in R^{s × 1}$). Considering this measure, ecological constraints often limit usable information ($U \subset Y_{1:N}$), creating the phenomenon of data-rich but information-poor environments \citep{bernus_data_2017,eichler_data_2019,forte_data_1994,ward_data-rich_1986}. From this framing, we qualitatively depict distinct vulnerability profiles across occupational contexts on this spectrum.

\section{Discussion}
\label{sec:Discussion}
Understanding and protecting decision-making under uncertainty in modern contexts requires a shift toward capturing how people process information in the presence of information-based threats. Here, we offer a comprehensive understanding that spans how people process, act upon, and propagate both erroneous and veridical information under conditions of uncertainty. We begin by describing the construct of cognitive security, describing the degree to which people are cognitive secure, at the human level, providing a needed theoretical construct situated between field-level, conceptual descriptions of cognitive security and operational measures. Stemming from this, this study advances the understanding of cognitive security by providing a computational framework grounded in human information processing theory. We demonstrate this modeling framework’s descriptive power across established cognitive heuristics and information-based threat phenomena, providing a foundation for understanding cognitive security that synthesizes influential factors into a comprehensive, empirically grounded theory. This quantitative grounding provides a need foundation for developing systematic interventions within the broader field.

\subsection{Theoretical Implications of the Cognitive Security Framework}
By providing a formal account of human information processing towards the formulation of judgments and decisions, this effort offers theoretical insights into how information-based threats influence human cognition and observable behavioral outcomes via a single-process modeling framework. In this framework, cognitive resources determine the expressiveness of alternative (i.e., hypothesis and action) spaces, aligning with the continuous nature of cognitive resources and brain function and supporting the advancement beyond traditional single versus dual-process model debates \citep{de_neys_dual-_2021}. Our model explains the illusory truth effect through a likelihood evaluation biased toward not being misled, evaluating observed information to be more likely true than false. Hence, we provide a computational account of why repeated exposure increases perceived truthfulness of statements initially judged to be less true (or even false) via the established truth bias. Additionally, we capture how memory and metacognitive processes influence the decision-making process, with the continued influence effect \citep{johnson_sources_1994,lewandowsky_misinformation_2012} understood as salient priors recalled in association with perceived information, consistent with theories about the role of perceived familiarity in producing illusory truth \citep{arkes_generality_1989}. Further, the regulatory role of metacognition in reducing truth bias \citep{jalbert_only_2023} suggests that individuals can actively re-evaluate the likelihood that observed statements are true, providing a mechanism for robustness to information-based threats. Reflecting this mechanism, Bayesian inference predicts no changes from priors if there is an equal chance of a statement being true or false. This unification suggests that these vulnerabilities to information-based threats are emergent from information processing.

Another example of how this framework provides new insights is in reference to the apparent paradox of sharing behavior misaligned with veracity discernment \citep{pennycook_shifting_2021}. The behavior is exemplified by investigations into older adults’ information sharing behavior: specifically, why older adults take the action to share erroneous information at a higher rate, despite being less susceptible in their judgments to information-based attacks such as consumer fraud \citep{ross_contrary_2014}. Our model explains this through differential risk-benefit evaluations, which are goal-oriented. Older adults prioritize social goals when using social media, such as connecting with others, over gaining information \citep{sims_information_2016}.  This goal prioritization alters the value function in our framework, where social engagement benefits from sharing potentially false information can outweigh accuracy concerns. Supporting this explanation, hesitancy to share erroneous information has been attributed to reputational risks \citep{altay_why_2022}. Our model formalizes this by demonstrating that misaligned veracity discernment and sharing behavior emerge when the benefit of social engagement outweighs accuracy concerns, even when individuals correctly identify information as false. As the field of cognitive security continues to identify novel behaviors, this framework should continue to be refined to capture such findings. 

Beyond these simulations, our framework reveals that there are multiple additional avenues through which information-based threats can compromise cognitive security. The most direct pathway involves substituting truthful information with repeated, coherent erroneous information, gradually shifting judgments toward erroneous conclusions. More sophisticated indirect attacks can distract and reduce cognitive resources by introducing irrelevant information, exploit affective responses to manipulate key risk and benefit valuations, or target perceived source credibility for veridical sources. Additionally, heuristics-based exploits can anchor individuals to specific hypothesis sets \citep{green_referendum_1998,kahneman_valuing_1992,tversky_judgment_1974}, creating cascading effects on decision-making performance. Notably, this framework is not limited to these implementations. In the case of naturalistic decision-making, action selection may instead be iteratively compared to a baseline `good enough' goal value rather than evaluating the entire action space \citep{klein_recognition-primed_1993}. Moreover, the emergence of patterns of cognitive resource allocations in experts may explain prototyping and the emergence of effective heuristics noted in naturalistic decision-making \citep{klein_naturalistic_2015}. Therefore, just as intuition can support improved decision-making in appropriate contexts, the structure and demands of the environment must be considered when assessing vulnerability to information-based threats.

\subsection{Context-Dependent Vulnerabilities}
Collectively, these results are consistent with the idea that information-based threats exploit underlying human cognitive processes, and effective countermeasures must account for how context modulates these vulnerabilities. We identify three factors that determine context-specific vulnerability: decision-maker expertise and strategy, team structure, and characteristics of the information environment. We note that characterizing vulnerabilities (and the framework's implementation) must account for variations in decision-making strategies across operational and environmental contexts. For instance, naturalistic decision-making (NDM) research demonstrates that proficient decision-makers (experts in performing their task) characteristically differ from novices under constraints \citep{klein_naturalistic_2008,lipshitz_taking_2001,mosier_judgment_2010,orasanu_reinvention_1993}. This subdomain of information processing literature, NDM has offered alternative models for describing how proficient operators make decisions \citep{klein_rapid_1986}, including the recognition-primed decision (RPD) model \citep{klein_recognition-primed_1993} and variants of the information processing model \citep{wickens_human_2015}. 

In NDM research, it is well established that expert operators rely heavily on long-term memory rather than executing an exhaustive choice evaluation, as novices might \citep{klein_rapid_1986,klein_characteristics_1995}. For instance, experienced emergency operators (e.g., firefighters and military officers) adopt similar strategies when facing time pressure, information uncertainty, and vague goals \citep{lipshitz_taking_2001}. Within our framework, experts operate with collapsed hypothesis sets, narrowed attention to environment-specific informational cues, and reduced action spaces for decision-making. While this processing strategy efficiently handles time and uncertainty constraints \citep{klein_naturalistic_2008}, we note that it may become a potential vulnerability when memory-stored patterns or recognition cues are taken advantage of by information-based attacks.

Another naturalistic factor, the role of teaming should also be considered. Teams represent a crucial operational structure in high-stakes environments \citep{mosier_judgment_2010}, exhibiting emergent characteristics that may both enhance and complicate cognitive security. Teams function as highly interdependent systems sharing goals and mutually influencing decision-making processes \citep{hollenbeck_multilevel_1995}. The interdependent nature of team decision-making creates both potential robustness and vulnerability to information-based threats. Both individual and team-level cognitive security are likely crucial mediators of team-level cognitive security when confronting information-based threats.

One factor not commonly considered, but which our framework elucidates as a critical aspect of cognitive security, is the impact the information environment has on cognitive security. In addition to the previously described naturalistic factors, the physical environment can influence decision-making \citep{wickens_engineering_2015} by limiting access to information at the beginning of human perception and a person’s cognitive resources. Such ecological constraints can restrict how much information can be utilized in a given environment \cite{petitet_computational_2021}. We capture these effects with our information processing spectrum (Figure \ref{fig:fig5}). Low information processing environments, characterized by limited usable information, are typically accompanied by team structures where cognitive attacks may interact with interpersonal social mechanisms \citep{grossman_teamwork_2017}. Conversely, high information processing environments present complex sourcing networks and impersonal social structures (such as individuals at home interacting with social media) \citep{subramanian_influence_2017}. We can hypothesize that individuals and teams in low information processing settings encounter and process cognitive attacks in a different manner than those who are abstracted from the environment with access to a multitude of sources of information.

In addition to these elements, extreme environments (a subset of low information processing environments) introduce additional considerations. For instance, hypoxia, sometimes encountered in mountaineering and within a pilot’s cabin, affects risk and benefit valuations \citep{pighin_decision_2012,pighin_loss_2014}. Isolated and confined, extreme (ICE) environments, such as spaceflight environments, exhibit altered psychological and social dynamics \citep{palinkas_psychosocial_2021} while requiring critical decisions under uncertainty. ICE environments pose a unique setting where ensuring sound decision-making, thus cognitive security, is paramount to mission success and operator safety. In contrast, attacks in high information processing environments may target more opaque outcomes associated with political affiliations and national sympathies \citep{allcott_social_2017}.

Another category involves transitional operators who encounter information-based threats in one environment that influence performance in another. Military drone pilots exemplify this category, operating in low information tactical environments while potentially influenced by high information social media exposure regarding the same operational contexts and facing unique psychological stressors \citep{saini_cry_2021}. Future research should examine how exposure to cognitive attacks from social media or news outlets spills over into operational decision-making.

\subsection{Observable Behavioral Outcomes of Cognitive Security}

Our theoretical analysis and modeling framework identified three observable behavioral outcomes that comprehensively evaluate cognitive security across the information processing chain. Veracity discernment reflects judgment formation, or the cognitive evaluation of information truthfulness; task-oriented actions capture decision implementation, or the translation of judgments into goal-directed behavior; and information sharing captures communication outputs, behaviors that propagate information and shape others' informational environments. These three outcomes represent distinct dimensions of information processing, yet they can dissociate from one another. For example, accurate veracity discernment does not guarantee appropriate task actions (such as when affective valuations are compromised), nor does it prevent the sharing of erroneous information (such as when social engagement goals dominate concerns of accuracy). By measuring all three outcomes, researchers can identify specific vulnerabilities within the information processing chain and develop targeted interventions. Central to these outcomes is their alignment with truth. While we acknowledge that people and organizations may define or interpret truth differently, our construct treats truth as an objective reference point. However, operational measures may vary depending on the specific epistemic or organizational anchor adopted.

Judgments are proxied through veracity discernment evaluations, which are the first observable behavioral outcome. Veracity discernment reflects how well queried judgments align with constructive, veridical sources of information. While existing literature has often focused on evaluating judgments of headlines or statements \citep{murphy_what_2023}, this construct should be understood more broadly. Veracity discernment encompasses how individuals comprehend their surrounding environment and evaluate the likelihood of outcomes (i.e., projection) \citep{endsley_combating_2018}. Therefore, veracity discernment should measure whether individuals' internal models of the world align with reality across these two core dimensions of situation awareness \citep{endsley_toward_1995}. Critically, veracity discernment can only be meaningfully assessed when constructive, veridical information is available within the environment, a prerequisite that also applies to the other behavioral aspects of cognitive security discussed below.

Several practical considerations apply to implementing veracity discernment as a behavioral measure. First, veracity discernment questions should be task-oriented and not aimed at assessing underlying beliefs and worldviews, which do not directly indicate task-specific judgment quality. Further, because assessing underlying truth judgments typically requires querying choice selection provided by experimenters (e.g., selecting ‘true’ or ‘false’ is a two-alternative forced-choice task), choices must have no associated value to assess judgments psychometrically using a veracity discernment task. In accordance with Prospect Theory, deviations from no-cost alternative choices may confound judgment assessments as biased decision assessments. Adhering to these principles, veracity discernment can be computed as an outcome to evaluate underlying truth judgments as a proxy measure.

The next behavioral outcome is a directly observable task-relevant aspect of decision-making: task-oriented actions, which are the outcomes of an individual’s policy used to achieve a specific task. Task actions are an underexamined aspect of cognitive security when considered within the information-based threats literature and provide an outcome variable to assess the influences of cognitive attacks on task-specific goals as the outcome of a choice rule or policy. The evaluation of task actions should assess the degree to which actions align with what veridical information sources indicate should be performed. For example, in a navigation-based task, where does a pilot fly their aircraft when given a false heading over the radio? 

The third behavioral outcome is the choice to propagate erroneous information within a task group. This subset of action is distinct in that it represents direct communication (contrasting task actions), which may, in turn, be used as an influential information source by others. One way sharing of erroneous information can be observed is through an evaluation of the degree of truthfulness in the information shared. Practically, evaluations of erroneous sharing should occur unprompted, as prompted questions may be confounded as prompted veracity discernments, altering the inherent benefit and risk value contributions associated with choosing to share erroneous information amongst others. Further, the sharing of erroneous information as an aspect of cognitive security should only be evaluated when it is not intended to mislead adversaries. Practically, information assessments should occur between allied groups, as information shared with adversaries may be strategic, an action not reflective of cognitive security. From these considerations, information quality rests on a spectrum spanning: sharing of erroneous information, not sharing erroneous or truthful information, and sharing truthful information. 

An important question raised by this framework concerns its applicability beyond operational or occupational settings, particularly to general population exposure to large-scale influence campaigns. Although such campaigns targeting the general public may constitute information-based threats that seek to erode trust and increase polarization, human cognitive security (as defined here) remains applicable to such contexts when individuals form incorrect judgments or take actions that do not align with constructive, veridical information about the world. Within this framework, beliefs, attitudes, trust, and ideological alignment, while potentially influential, are not direct outcomes of cognitive security. Cognitive security can only be empirically measured in such contexts when judgments, actions, or information sharing can be evaluated relative to veridical information.

Importantly, a number of potential correlates (i.e., indicators) may be assessed that do not directly reflect a person or team’s cognitive security, including beliefs, attitudes, and engagement behaviors.  These secondary outcomes may be useful as indicators of cognitive security rather than as valid construct measures for quantifying the degree to which a person is cognitively secure. Beliefs (and worldviews) are expected to influence priors \citep{zmigrod_misinformation_2023} and the set of hypotheses considered, which in turn may affect task-oriented judgments and decisions. Attitudes, such as favorability, can be expected to modulate risk and benefit evaluations for making decisions \citep{finucane_affect_2000}. Engagement behaviors, such as liking a post with erroneous information (utilized in recent experimental tasks (\citep{butler_misinformation_2024}) or clicking on a link that contains false information (or conversely, a fact check), may indicate heightened or decreased vulnerability via increased exposure to information-based threats or constructive veridical information, respectively. However, engagement with erroneous information does not directly indicate decreased cognitive security, nor does engaging with truthful information indicate increased cognitive security. Supporting this distinction, recent work examining belief-confirming and belief-examining click behavior (i.e., engagement with information that confirms and contradicts priors, respectively) found that increasing engagement with veridical belief-examining information does not lead to better veracity discernment on its own \citep{tanaka_beyond_2025}.

\subsection{Cognitive Security Profiles}
To advance empirical research in this field, our three-outcome framework distinguishes between primary outcomes (veracity discernment, task-oriented actions, and information sharing) and secondary indicators (beliefs, attitudes, engagement behaviors). Using these primary outcomes in conjunction, we can characterize a fully cognitively secure individual as impervious to erroneous and misleading information when discernible truthful information sources are also available (i.e., situations characterized by conflicting information). As a result, their judgments correctly reflect constructive veridical information, their actions align with what constructive veridical information sources indicate they should do, and they do not propagate erroneous information as truth within their organization. In contrast, a fully compromised individual utilizes erroneous or misleading information to make judgments; their actions align with what erroneous or misleading information sources indicate they should do; and they propagate erroneous information as truth. While the aspects presented thus far are framed on the level of individuals, these measures can capture team vulnerability to cognitive attack by observing team-level task-associated actions.  

Our framework's mechanistic insights reveal multiple intervention pathways for enhancing cognitive security. Interventions may target resource allocation (such as through metacognitive prompting or procedural scaffolding), source credibility assessment (such as through prebunking or calibrated trust-building), or affective valuation (such as through emotional regulation strategies). Each pathway addresses distinct vulnerabilities: resource allocation and credibility assessment interventions improve the judgments that mediate behavior, while affective valuation interventions directly influence decision-making. Prebunking and debunking approaches show varying effectiveness against erroneous claims \citep{brashier_timing_2021,jolley_prevention_2017,tanaka_beyond_2025,tay_comparison_2022,vraga_testing_2020,walter_how_2018}, reinforcing the context-dependent nature of these interventions. Therefore, systematic evaluations of how these intervention mechanisms perform across different operational contexts and information environments remain critical \citep{crum_misinformation_2024}. Moreover, understanding the neurocognitive mechanisms that underlie human cognitive security may lend to the development of protective AI-based tools for intervention development \citep{crum_understanding_2025}.

\subsection{Limitations and Future Work}
Several important limitations constrain the current modeling framework and point toward critical avenues for future research. Primarily, the presented framework incorporates constructs pertaining to cognitive resources and affective valuation of outcomes to characterize the decision-making on an individual level. However, both elements are multidimensional. Cognitive resources, for instance, have been described in relation to cognitive processes such as attention and memory, as well as energetic elements such as mental workload and neurobiological resources such as brain oxygenation. The multidimensional nature of resources \citep{wickens_multiple_2002,wickens_multiple_2008}, and often abstract descriptions, have left the construct difficult to operationalize \citep{navon_resourcestheoretical_1984}. Here we provide an alternative, needed operationalization of cognitive resources \citep{hommel_editorial_2022} as an allocation of resources along a domain of hypotheses, mirroring how neurophysiological mechanisms can lead to the prioritization of specific hypotheses and can curtail the consideration of alternatives \citep{dehais_momentary_2019,dehais_neuroergonomics_2020}. However, this representation does not yet fully capture contributing factors or their interactions. Similarly, we model the influence of affect as the end-product of an outcome’s affective valuation. However, affect too is multifaceted, including degree of valence, arousal, and specific emotional categories \citep{kollias_affect_2021,kuppens_relation_2013,posner_circumplex_2005,russell_circumplex_1980}. 
Future empirical examinations should seek to parse out the influence of these constructs’ subdimensions to develop effective countermeasures aimed at preserving cognitive security across a broad range of scenarios. Regarding cognitive processes such as memory and attention, incorporating our modeling framework into existing cognitive architectures, such as the Adaptive Control of Thought-Rational (ACT-R) \citep{anderson_ensuring_2024}, may provide additional useful predictive insights for memory-based phenomena. To date, ACT-R has been used to formalize and examine an erroneous information phenomenon not examined here: the continued influence effect \citep{hough_model_2024,hough_modeling_2024}, suggesting promising avenues of integration. Regarding affective influences, future work should also distinguish the effects of both task-related affect (emotions directly tied to the information processing task) and task-adjacent affect (background emotional states that may influence processing) \citep{mosier_role_2010}. Both types of affect across multiple dimensions (including arousal, valence, and specific emotional categories) should be studied systematically, particularly regarding their differential influence on outcome valuation and information processing strategies. Finally, future research should address how these multidimensional constructs interact across different temporal scales and contextual factors, as information-based vulnerabilities may depend on the specific combination of resource limitations and affective states present at the time of exposure.

\section{Conclusion}
\label{sec:Conclusion}
The expansion of sophisticated information-based threats requires a fundamental advancement within the field of cognitive security. Here, we introduce a human cognitive security construct that bridges field-level definitions with operational measures by characterizing the degree to which people are cognitively secure within the context of their judgments and decisions. From this construct, we develop an integrative modeling framework that synthesizes insights from information-based threats and the broader information processing domain into a unified computational structure. By integrating Bayesian inference with Prospect Theory, we demonstrate how information-based threats systematically exploit fundamental cognitive mechanisms rather than representing unique vulnerabilities. The framework captures established canonical phenomena, including the illusory truth effect and misaligned veracity discernment and sharing behavior, while revealing how cognitive resource allocation, affective valuation, and source credibility jointly shape vulnerability. This framework generates three observable behavioral outcomes for measuring cognitive security: veracity discernment, task-oriented actions, and information sharing, each grounded in a specific dimension of information processing. Our information processing spectrum illustrates how both data availability and ecological constraints determine cognitive security across operational contexts, from isolated and confined extreme environments to high-information social media settings. As information environments continue to evolve, this framework provides the theoretical foundation needed to develop, test, and deploy effective countermeasures that protect judgments and decision-making under uncertainty.

\section{Methods}
\label{sec:Methods}

\subsection{Framework Architecture}

Our developed framework models the behavioral aspects of cognitive security (Figure \ref{fig:fig1}b) as outcomes of a process through which cognitive resources and affective valuations modulate judgments and decisions (Figure \ref{fig:fig1}c; see detailed Methods). This framework captures how cognitive, affective, and social factors mediate the behavioral aspects of cognitive security. The architecture capturing the central processing of information is framed through a descriptive Bayesian inference lens on the front end, where an iterative weighing of cumulative evidence from incoming information with priors from past experiences forms new assessments for making judgments and decisions. On the back end, considered actions are evaluated via Prospect Theory.

\subsection{Cognitive Encoding of Information}

To quantitatively define cognitive processing of information in the environment, we adapt a resource mapping model used in perceptual Bayesian observer modeling \citep{hahn_unifying_2024,wei_bayesian_2015} to traverse from a physical, uncertain information cue to a cognitive likelihood assessment denoting comprehension. In the perceptual domain, finite coding resources ($S$) are allocated across stimuli $\theta \in X$,
\[
S = \int_\theta R(\theta) d\theta \le C ,
\]
where $R(\theta)$, the neural coding of resources, is related to the mapping of a one-dimensional variable physical stimulus to an abstracted neural representation ($m \in Y$) encoded by a population of neurons affected by noise. The mapping $F:X \to Y$, from stimulus space to neural space, and additive noise ($\eta$) is represented as follows:
\[
m = F(\theta) + \eta, \quad R(\theta) \propto F'(\theta).
\]

From a chosen mapping, the true (i.e., generative) likelihood of a stimulus given a given measurement, $y$, within a considered stimulus space (necessary for inference in stimulus space), is the distribution of $F^{-1}(y)$:  
\[
L(\theta; y) = p_X(y \mid \theta) \propto p_Y(y \mid F(\theta)) \cdot R(\theta).
\]

Considering a stimulus with uncertainty (i.e., a noisy realization, $\tilde{\theta}$, as opposed to a draw from a stimulus distribution), as a hierarchical generative model:
\[
L(\theta; y) = p_X(y \mid \theta) \propto \int R(\tilde{\theta}) p_Y(y \mid F(\tilde{\theta})) \cdot p(\tilde{\theta} \mid \theta) \, d\tilde{\theta}.
\]

Thus, the likelihood function utilized by the central nervous system can be expressed as a mapping $\Phi_R: p(\tilde{\theta} \mid \theta) \mapsto L(\theta; y)$ from the underlying stimulus to a measurement-dependent distribution over possible stimulus values, shaped by the resource-constrained neural representation, such that:
\[
L(\theta; y) = [\Phi_R p(\tilde{\theta} \mid \theta)](y).
\]

We extend this representation from perceptual to cognitive information processing by replacing the stimulus space with a finite hypothesis space, $H$. Unlike perceptual processing, where $L(\theta; y)$ is tied to a physical stimulus, cognitive processing involves evaluating the likelihood of candidate hypotheses \citep{tenenbaum_theory-based_2006}, analogous to the diagnosis and reliability assessment of cues \citep{ wickens_engineering_2015,wickens_information_2021}. Because the true form of a likelihood function is unknown (i.e., whether it actually follows the generative likelihood) and because the hypothesis space does not necessarily correspond to a stimulus space, we define a generalized form of the neural likelihood:
\[
L(H; y) = R_H [p(\tilde{x} \mid x)](y).
\]

Here, the likelihood function is represented as the output of a mapping $R_H: p(\tilde{x} \mid x) \mapsto L(H; y)$ between the information source outside of human information processing $L(x; \tilde{x}) = p(\tilde{x} \mid x)$ and the cognitive resources allocated along a finite hypothesis space, maintaining that cognitive resources are finite. During our simulations, we explore how different cognitive resource mappings result in different judgments and actions.

This formulation provides the additional utility of considering an upper bound of the information available to be processed by the human observer. Consider the case of an ideal observer, free of the ecological, computational, and noise constraints of the human, such that a finite set of measurements in the environment, $Y_{1:N}$, are processed with infinite computational resources, and perfect measurement (i.e., no sensory noise) occurs. A noisy stimulus characterized by the generative likelihood function becomes $L(x; Y_{1:N}) \approx L(x; \tilde{x})$. Thus, the upper bound on the information available to a human within the environment, using an ideal observer, can be characterized by the Fisher Information:
\[
J(x) = \mathbb{E}\left[\left(\frac{\partial}{\partial x} \log L(x; Y_{1:N}) \right)^2 \right].
\]

\subsection{Judgment Formulation}

Following the cognitive encoding and comprehension of external information, a decoding probabilistic assessment is formulated according to Bayes' rule:
\[
p(H \mid y) \propto p(y \mid H) p(H).
\]

This probabilistic assessment is equivalent to the judgment of a hypothesis given a prior $p(H)$. Similar to the likelihood function, the prior is influenced by the reasoner’s background knowledge \citep{tenenbaum_theory-based_2006} and worldview \citep{cook_rational_2016, zmigrod_misinformation_2023}.

Evidence evaluation in our framework depends on four primary factors: information consideration, a credibility assessment, the variability of the underlying information (i.e., cue uncertainty), and the amount of evidence evaluated (inversely proportional to measurement uncertainty). Information consideration captures what information in the environment is attended to, processed, and ultimately considered, as cues must be perceived and require cognitive resources \citep{wickens_information_2021}. Perceived information credibility alters likelihood evaluations. For instance, statements (given by incoming information, $Y$) from sources deemed credible are evaluated as more true than false $[p(Y \mid H:\text{true}) > p(Y \mid H:\text{false})]$, while statements from sources deemed non-credible are evaluated as more false than true $[p(Y \mid H:\text{false}) > p(Y \mid H:\text{true})]$. Importantly, this mechanism operates through evidence evaluation rather than prior reliance alone. Cue uncertainty reflects the inherent variability of underlying information and influences how strongly evidence hypotheses are weighted. The amount of evidence plays a role through prior probability updates from posterior estimates. Short and long-term memory, deliberation time, and metacognition all feed into this resource and affect modulated information processing model, qualitatively depicted in Figure \ref{fig:fig1}c.

In our simulations demonstrating judgment heuristics, we use an arbitrary prior. In our simulations of veracity discernment and sharing behavior, we use a uniform prior to simulate naivety to unseen (no prior knowledge), cold (no interaction with worldview) media. Both choices enable evaluations of how cognitive resources and affective valuation influence judgments and decision-making.

\subsection{Action-Based Outcome Value Evaluation}

From the posterior distribution $p(H \mid y)$, probabilistic assessments of future outcomes, associated with various decisions, are then cast as $p(O \mid y)$. This process is analogous to a mapping from Level 2 situation awareness (i.e., comprehension) to Level 3 situation awareness (i.e., projection) \citep{endsley_toward_1995}. In the case of considering self-action (vs. the case of considering `no action'), this process is equivalent to the mental simulation of action \citep{klein_recognition-primed_1993}.

Conveniently, our simulations enable a simplification of this mapping by assuming that the projection mapping from $P: p(H \mid y) \mapsto p(O_\text{correct} \mid y)$ is identity in the case of selecting the correct answer during a veracity discernment task. For example, consider the judgment that a statement is true $p(H=6 \mid y)=0.8$. In this scenario, we assume that the probability of selecting the correct choice when taking the action `select 6' on self-report is equivalent to the posterior of that hypothesis, thus $p(O_\text{correct} \mid y)=0.8$.

In the presence of outcome probabilities and associated values, we utilize Cumulative Prospect Theory \citep{tversky_advances_1992} in this mathematical framework. Prospect theory originally accounts for deviations from Expected Value theory, capturing how individuals make decisions under uncertainty by accounting for reference dependence, loss aversion, diminishing sensitivity, and the overweighing of small probabilities, and the overweighing of large probabilities \citep{kahneman_prospect_1979}. Supporting more complex scenarios, we utilize the following Cumulative Prospect Theory action valuation formula in discrete form:
\[
V_A = \sum_{i \in \text{gains}} \pi_i^+ \cdot v(O_i) + \sum_{j \in \text{losses}} \pi_j^- \cdot v(O_j).
\]

Here, the prospective value of an action, $V_A$, depends on the associated outcomes considered and these outcomes’ associated values. In accordance with Cumulative Prospect Theory, the design weights, $\pi$, transform the cumulative probabilities of outcomes, depending on if the outcome has an associated loss or gain, using probability weighting functions. In these simulations, we use the weighting function introduced by \citet{tversky_advances_1992}, with canonical weighting curvature parameters for gains and losses ($\gamma^+ = 0.61$ and $\gamma^- = 0.69$ respectively). Further, the value functions, $v(O)$, transform values such that gains are concave ($\alpha=0.88$), losses are convex ($\beta=0.88$), and losses loom larger than gains ($\lambda=2.25$).

In Cumulative Prospect Theory, these value functions transform values provided to humans during experiments. In our formulation, these functions alter affect and goal-modulated values, which we refer to as affective valuation. We model affective valuation given the affect heuristic \citep{alhakami_psychological_1994, finucane_affect_2000}, where risks ($\pi_j^- \cdot v(O_j)$) and benefit ($\pi_i^+ \cdot v(O_i)$) appraisals are modulated by affect. This approach is consistent with more recent works that have sought to model how affect changes the value function across various experimental paradigms \citep{lerner_emotion_2015,phelps_emotion_2014,schulreich_fear-induced_2020}. Within this framework, we explore how affective valuation influences decisions in two simulations: modeling of the affect heuristic and modeling discrepant veracity discernment and sharing behavior.

\subsection{Action Determination}

Consistent with previous implementations of action determination, we utilize the distribution of prospective values over considered actions to inform action selection. In the case of a continuous (e.g., selecting a truth rating value on a scale between 1 and 6) or ordinal action space, we model action selection as the action that minimizes the mean squared error (MSE) cost function, corresponding to the mean of the prospective value function across the action space. This approach enables a direct demonstration of how factors influence alternative action values on average. Previous sensorimotor efforts have utilized MSE across continuous domains \citep{kording_bayesian_2014,kording_bayesian_2004}. In nominal action cases (e.g., share or do not share), we implement a maximize expected gain cost function (i.e., a greedy decision maker), such that the chosen action maximizes the expected gain across all considered actions. The cost function utilized for decision-making is likely context-specific and remains an open question in decision-making sciences \citep{tassinari_combining_2006}.

In our simulations, we constrain the action spaces by simulating simple alternative choice paradigms. In the case of veracity discernment, the action space $A$ is constrained to the continuum provided to participants in an analogous experimental paradigm \citep{hassan_effects_2021}, such that $A \in [1,6]$, where 1 corresponds to `not truthful' and 6 corresponds to `very truthful.' In the case of information sharing, the action space is constrained to two discrete actions: share and do not share.

This modeling framework also enables capturing probabilistic decision strategies for both continuous and categorical action spaces. The prospective value of each alternative can be converted into a probability density from which actions are drawn using the Luce-Shepard choice rule \citep{luce_individual_1959,shepard_stimulus_1957}, which functions as a softmax function across possible actions. For comparison to the mean empirical illusory truth data, we also considered the mean of this distribution,
\[
\mu_A = \mathbb{E}\left[\frac{e^{\beta V_A}}{\int_A e^{\beta \bar{V}_A} \, d\bar{A}}\right],
\]
which can similarly capture the illusory truth data (MSE = 0.015; $R^2 = 0.89$) with the introduction of a free parameter, $\beta$, which determines how deterministically a decision maker behaves.

\section{Acknowledgments}
The authors acknowledge the Air Force Office of Scientific Research (AFOSR), award no. FA9550-23-1-0453, for support of this research. The views expressed in this report are those of the authors and do not reflect the official policy or position of the US Air Force, Department of Defense, or the U.S. Government.

\section{Author Contributions}
A.R.A. conceived the study. 
A.R.A., E.E.R., and A.P.A.H. developed the theoretical framework. 
S.R.B, R.E.N, and J.C. provided critical ideas and feedback that shaped the research direction. A.R.A. analyzed the data. 
L.H, A.P.A.H., R.E.N, and C.T. led funding acquisition and project supervision. 
A.R.A., A.P.A.H., and E.E.R. interpreted the results and prepared the initial manuscript. 
All authors contributed to reviewing and editing the manuscript. 
All authors approved the final version for publication.

\printbibliography






\end{document}